\documentclass{epl}

\title{Roughness of tensile crack fronts in heterogenous materials}
\shorttitle{Roughness of crack fronts}
\author{E. Katzav \inst{1} \and M. Adda--Bedia \inst{1}}
\institute{
  \inst{1} Laboratoire de Physique Statistique de l'Ecole Normale Sup\'erieure,
24 rue Lhomond, 75231 Paris Cedex 05, France.}


\begin{document}

\maketitle

\begin{abstract}
The dynamics of planar crack fronts in heterogeneous media is
studied using a recently proposed stochastic equation of motion that
takes into account nonlinear effects. The analysis is carried for a
moving front in the quasi-static regime using the Self Consistent
Expansion. A continuous dynamical phase transition between a flat
phase and a dynamically rough phase, with a roughness exponent
$\zeta=1/2$, is found. The rough phase becomes possible due to the
destabilization of the linear modes by the nonlinear terms. Taking
into account the irreversibility of the crack propagation, we infer
that the roughness exponent found in experiments might become
history-dependent, and so our result gives a lower bound for
$\zeta$.
\end{abstract}

\section{Introduction}

The dynamics of cracks in heterogeneous media is a rich field
encompassing a large range of physical phenomena. In such situations
a commonly studied quantity is the so called roughness exponent
$\zeta$. However, it is important to distinguish at least three
different roughness exponents \cite{bouchaud}: one describing the
roughness in the direction perpendicular to the crack propagation, a
second the roughness in the direction of the propagation, and a
third one (which will interest us in the following) describing the
in-plane roughness of the crack front during its propagation through
the material. The exponent characterizing this in-plane roughness,
has been measured in different materials, where it was found to be
in the range $0.5$--$0.6$, over at least two decades
\cite{Daguier,Schmittbuhl}. Despite numerous efforts
\cite{Rice,Ramanathan97,Schmittbuhl03}, there is unfortunately no
satisfactory theory that predicts the value of this exponent.

In this paper we intend to contribute to the theoretical
understanding of this problem. For that purpose we use an equation
of motion of the crack front $h(x)$ derived previously \cite{prl05}.
This equation contains two important ingredients - irreversibility
of the propagation of the crack front and nonlinear effects. It is
given by
\begin{equation}
\frac{{\partial h}}{{\partial t}}\left( {x,t} \right) = \sqrt {1 +
h'^2 } \left[ {K_I  \left( h \right) - K_c \left( {x,h} \right)}
\right]\Theta\left[ {K_I - K_c}
\right] \;,\label{eq:start}
\end{equation}
where $h'=\partial h/\partial x$, $\Theta(\cdot)$ is the Heaviside
function, $K_I(h)$ is the stress intensity factor of the crack front
(calculated to second order in $h$ \cite{prl05}) and $K_c \left(
{x,h} \right)$ is a random term representing the heterogeneity in
the local material toughness due to disorder. The random term can
always be separated as $K_c \left( {x,h} \right) = K^*  + \eta
\left( {x,h} \right)$, where $K^*$ is an average toughness and
$\eta$ is its fluctuating part. It should be emphasized that terms
including the effects of the system size were consistently left out
in the derivation of this equation \cite{prl05}.

Solutions of stochastic growth models such as Eq.~(\ref{eq:start})
exhibit scaling behavior which is described using the time dependent
height-height correlation function
\begin{equation}
\left\langle \left[ h(x,t) - h(x',t') \right]^2 \right\rangle =
|x-x'|^{2\zeta} f\left( \frac{|x-x'|}{|t-t'|^{z}} \right)
\label{scaling},
\end{equation}
where $\zeta$ is the roughness exponent of the interface and $z$ is
the dynamic exponent. The brackets $\left\langle
\cdots\right\rangle$ denote average over disorder. In the following,
we discuss several observations and arguments in order to explain
our approach. This will help to simplify the equation of motion and
to apply the self-consistent-expansion (SCE) approach in order to
derive results for the scaling exponents. This method was developed
by Schwartz and Edwards \cite{SCE,TCN04} and has been applied
successfully to the Kardar Parisi-Zhang (KPZ) equation \cite{KPZ}.
The method gained much credit by being able to give sensible
predictions for the KPZ scaling exponents in the strong-coupling
phase above one dimension where many renormalization group (RG)
approaches failed \cite{Wiese98}. Another point which is especially
relevant for our purpose is that for a family of models with
long-range interactions (of the kind treated presently) SCE
reproduced exact one-dimensional results while RG failed to do so
\cite{Nonlocal}.

\section{The general approach}
Regarding Eq.~(\ref{eq:start}), we expect to see three different
regimes: A static regime for which $K_0 \ll K^*$ (where the
Heaviside function in (\ref{eq:start}) can be safely approximated by
$0$); A regularly moving interface for large values of $K_0$ (where
the Heaviside function can be safely approximated by $1$); And an
intermediate complex regime, where $K_0 \sim K^*$. In this last
regime, a very important factor seems to be the stabilizing terms
which were dropped out in the derivation \cite{prl05}, but which
will make sure that the crack will stop after a while (as indeed
seen in experiments \cite{Daguier,Schmittbuhl}).

Based on that picture, we hypothesize that a frozen dynamically rough interface is seen in experiments \cite{Schmittbuhl}), rather than a rough phase determined by a static pinned interface. In other words, we stress the point that the crack tends to stop due to its physical nature even without the presence of heterogeneities. This is indeed the case in cantilever beam experiments \cite{Schmittbuhl}, where the crack faces are increasingly opened in order to induce crack front motion. As a result the front starts moving until it stops. The heterogeneities only induce roughness and as we argue, a dynamical roughness, which is then frozen due to the irreversibility of the fracture process.

In order to test this picture, we approximate this system by neglecting consistently all mechanisms which deal with the slowing down of the interface, as well as the freezing of it. The assumption here is that the specific aspect of fine-tuning the opening stress mode (for example by imposing a time-dependent external loading) is exactly what the experiment does. Then we analyze the system at that critical point whichever means were taken to get there. This involves neglecting the Heaviside function on the right-hand side of Eq.~(\ref{eq:start}). We suspect that this term does play a role in the final stages of freezing, namely by imposing differential arrest along the interface (note again that the interface would stop anyway, even without this term). This would tend to increase the roughness. Thus, we would consider the
results obtained below as a lower bound for the roughness, offering a quantitative physical explanation up to the last steps of the freezing.

Interestingly, a similar approximation is implicitly present in the KPZ system. It is well known that any rough surface would eventually flatten by the KPZ system if the noise is stopped \cite{EytanBD}. However in real situations, it is compensated by non-zero ``angle of repose'' that eventually freezes the system in a rough phase \cite{Moshe04} (this is expressed by an additional Heaviside function in the KPZ equation). It is also shown that the roughness exponent would be the same as that of the driven system if the freezing is done adiabatically \cite{Moshe04}. This shows that a hysteretic effect (existence of an angle of repose) can be consistently neglected once an above-threshold driving noise is present. It also hints that the final roughness of our system would be ``history dependent" i.e. it would depend on the protocol of the loading/freezing, if not done adiabatically.

Following the previous arguments, we approximate the noise term for
the moving front, where $h \simeq vt$, by $\eta \left( {x,h}
\right)\simeq \eta \left( {x,vt} \right) = \hat \eta \left(
{x,t}\right)$ \cite{LeDoussal04}. Also, we do keep nonlinear terms,
since we claim (and will justify later) that they play an important
role in roughening the interface. Obviously a linear equation of the
kind described above (i.e. taking into consideration only the linear
term in $K_I(h)$) would not yield any roughness, and actually even
if the KPZ nonlinearities (i.e. $h'^2$ terms) are kept, we would
also end up with a smooth surface, or at most logarithmically rough
(this is a special case of the so called Fractal KPZ equation
studied previously in \cite{FKPZ03}). When keeping consistently
second order terms, the resulting equation of motion becomes
\begin{eqnarray}
 \frac{{\partial h}}{{\partial t}} &=&  K_0 \int\limits_{ - \infty }^\infty  {\frac{{h'\left( {x'} \right) }}{{\left( {x' - x} \right) }}\frac{{dx'}}{{2\pi }}}  + K_0 \int\limits_{ - \infty }^\infty  {\frac{{dx'}}{{2\pi }}\int\limits_{ - \infty }^\infty  {\frac{{dx''}}{{2\pi }}\frac{{{h'\left( {x'} \right) } {h'\left( {x''} \right) }}}{{\left( {x' - x} \right) \left( {x'' - x'} \right) }}} } \nonumber \\
 &-& \frac{3}{8}\left( {\frac{4}{3}K^*  - K_0 } \right)h'^2 +\left( {K_0  - K^* } \right)  + \hat \eta \left( {x,t} \right)\;,
\label{eq:motion}
\end{eqnarray}
with noise correlations described by
\begin{equation}
\left\langle {\hat \eta \left( {z,t} \right)\hat \eta \left( {z',t'}
\right)} \right\rangle  = 2D\delta \left( {z - z'} \right)\delta
\left( {t - t'} \right)\;,
\end{equation}
where $D$ is the variance of the noise. The integrals in
Eq.~(\ref{eq:motion}) should be taken in the sense of Cauchy
principal value. The first prediction of this equation is that the
mean velocity should be proportional to the constant term, i.e. $v
\propto (K_0  - K^*)$. The natural appearance of a velocity
strengthens the simplification of taking a time-dependent noise
term. For convenience, the constant term can be scaled out by
transforming into a co-moving coordinate system, i.e. $h \rightarrow
h+(K_0-K^*)t$. Then, by looking at the KPZ term (i.e. $h'^2$) we can
estimate the region where this discussion is relevant. Roughly, when
the coefficient of that term remains negative (i.e. for $K_0 <
{\textstyle{4 \over 3}}K^*$), we are still in the quasi-static
regime since in that case a rough interface would decrease the
velocity, while for higher values of the applied stress ($K_0  >
{\textstyle{4 \over 3}}K^*$) the system would be in the regularly
moving regime. This estimate is consistent with our assumption that
the dynamics of interest is not necessarily at $K_0\simeq K^*$, but
in some range above it (i.e. $K^* \leq K_0 \leq \frac{4}{3}K^*$). In
the following, we will neglect the KPZ-term to simplify the
presentation. We checked that the results we get for the scaling
exponents do not depend on its existence (beyond its stabilizing
effect). At the end, we will comment on the modifications due to its
presence.

\section{The SCE method}
The SCE method is based on going over from the Fourier transform of the equation in Langevin form to a Fokker-Planck form and on constructing a self-consistent expansion of the distribution of the field concerned. We then consider the simplified version of the equation of motion in Fourier components
\begin{equation}
\frac{{\partial h_q \left( t \right)}}{{\partial t}} =  - c_q h_q  -
\sum\limits_{\ell ,m} {M_{q\ell m} h_\ell  h_m }  + \hat \eta _q
\left( t \right)
\label{eq:motionFourier},
\end{equation}
where $c_q  = \frac{{K_0 }}{2}\left| q \right|$ and $M_{q\ell m}  =
- \frac{{K_0 }}{{4\sqrt{L}  }}\left| q \right|\left| \ell
\right|\delta _{q,\ell  + m}$, $L$ being the size of the front. Note that in contrast to the KPZ
problem $M_{q\ell m}$ has the symmetries $M_{q\ell m} =
M_{-q,\ell,m} = M_{q,-\ell,m} = M_{q,\ell,-m}$. Last, $\hat
\eta_q(t)$ is a noise term with zero average described by its
variance $\left\langle {\hat \eta _q \left( t \right)\hat \eta _{q'}
\left( {t'} \right)} \right\rangle  = 2D\delta \left( {q + q'}
\right)\delta \left( {t - t'} \right)$. Rewriting this equation in a
Fokker-Planck form we get
\begin{equation}
\frac{{\partial P}}{{\partial t}} + \sum\limits_q {\frac{\partial
}{{\partial h_q }}\left[ {D_0 \frac{\partial }{{\partial h_{ - q} }}
+ c_q h_q  + \sum\limits_{\ell ,m} {M_{q\ell m} h_\ell  h_m } }
\right]P}  = 0\;,
\label{eq:motionFokker}
\end{equation}
where $P\left( {\left\{ {h_q } \right\},t} \right)$ is the
probability functional for having a height configuration $\left\{
{h_q } \right\}$ at time $t$.

The expansion is formulated in terms of the steady-state structure
factor $\phi_q = \left\langle {h_{-q} h_q} \right\rangle$ (or
two-point function), and its corresponding steady-state decay rate
that describes the rate of decay of a disturbance of wave vector $q$
in steady state, namely $\omega_q^{-1} = \left.{\int_0^\infty
\left\langle {h_{-q}(0) h_q(t)} \right\rangle dt} \right/
\left\langle {h_{-q} h_q } \right\rangle$. From the scaling form
(\ref{scaling}) it follows that for small $q$'s $\phi_q$ and
$\omega_q$ behave as power laws in $q$, namely $\phi_q =
A|q|^{-\Gamma}$ and $\omega_q = B|q|^z$, where $z$ is the dynamic
exponent, and the exponent $\Gamma$ is related to the roughness
exponent by $\zeta=(\Gamma-1)/2$.

The main idea of SCE is to write the Fokker-Planck equation $\partial
P/\partial t= OP$ in the form $\partial P/\partial t= \left[O_0 + O_1 + O_2 \right]P$, where $O_0$, $O_1$ and $O_2$ are zero, first and second order operators in some parameter. The evolution operator $O_0$ is chosen to have a simple form $O_0  =  - \sum\limits_q {\frac{\partial }{{\partial
h_q }}\left( {D_q \frac{\partial }{{\partial h_{ - q} }} + \omega_q h_q } \right)}$, where $D_q/\omega_q=\phi_q$. Note that $\phi_q$ and $\omega_q$ are still unknown. Next, an equation
for the two-point function is obtained. The expansion has the form
$\phi_q = \phi_q + e_q \left\{ {\phi_p ,\omega_p } \right\}$, where $e_q$ is a functional of all $\phi$'s and $\omega$'s. This reflects the fact that the lowest order in the expansion is
exactly the unknown $\phi_q$. In the same way, an expansion for
$\omega_q$ is given by $\omega_q = \omega_q + d_q
\left\{{\phi_p ,\omega_p} \right\}$. Now, the two-point function
and the characteristic frequency are determined by setting
$e_q \left\{ {\phi _p ,\omega _p } \right\} = 0$ and $d_q \left\{
{\phi _p ,\omega _p } \right\} = 0$. To second order in
the expansion, we get the following two coupled integral equations
\begin{eqnarray}
D_0  &-& \frac{{K_0 }}{2}\left| q \right|\phi _q  + I_1 \left( q
\right)\phi _q  + I_2 \left( q \right) = 0 \;,\label{eq:Static2} \\
\omega _q  &-& \frac{{K_0 }}{2}\left| q \right| + J\left( q \right)
= 0 \;,\label{eq:Herring2}
\end{eqnarray}
with
\begin{eqnarray}
 I_1 \left( q \right) &=& \frac{{K_0^2 }}{32\pi}\left| q \right|\int {d \ell \left| \ell  \right|{\textstyle{{ \left| \ell  \right|\left( {\left| {q - \ell } \right| + \left| q \right|} \right) \phi _{q - \ell }+   \left| {q - \ell } \right|\left( {\left| \ell  \right| + \left| q \right|} \right)\phi _\ell} \over {\omega _q  + \omega _\ell   + \omega _{q - \ell } }}}} \;,\label{eq:integrals1a} \\
 I_2 \left( q \right) &=& \frac{{K_0^2 }}{32\pi}q^2 \int {d \ell \left| \ell  \right|{\textstyle{{\left( {\left| \ell  \right| + \left| {q - \ell } \right|} \right)\phi _\ell  \phi _{q - \ell } } \over {\omega _q  + \omega _\ell   + \omega _{q - \ell } }}}}  \;,\label{eq:integrals1b}\\
 J\left( q \right) &=& \frac{{K_0^2 }}{32\pi}\left| q \right|\int {d \ell \left| \ell  \right|{\textstyle{{\left| \ell  \right|\left( {\left| {q - \ell } \right| + \left| q \right|} \right) \phi _{q - \ell } +   \left| {q - \ell } \right|\left( {\left| \ell  \right| + \left| q \right|} \right)\phi _\ell} \over {\omega _\ell   + \omega _{q - \ell } }}}}\;.
\label{eq:integrals1c}
\end{eqnarray}
It is interesting to mention here that Eq.~(\ref{eq:Static2}) can be
understood as emanating from the short time balance of the original
equation, while Eq.~(\ref{eq:Herring2}) comes from its long time
balance \cite{TCN04}.

These equations can be solved exactly in the asymptotic limit
(i.e. for small $q$'s) to yield the required scaling exponents
governing the steady-state behavior and the time evolution. The
difficulty here arises from the fact that the integrals involved,
$I_1(q)$, $I_2(q)$, and $J(q)$, have contributions from large
$\ell$'s as well as from small $\ell$'s. Therefore, one must consider the contribution of the large $\ell$
integration domain on the small $q$ behavior of the integrals
(\ref{eq:integrals1a}-\ref{eq:integrals1c}). For this, we break up the integrals $I_i(q)$ and
$J(q)$ into the sum of two contributions $I_i^<(q)+I_i^>(q)$,
and $J^<(q)+J^>(q)$, corresponding to domains of $\ell$
integration with low and high momentum respectively. We expand
$I_i^>(q)$ and $J^>(q)$ for small $q$'s and obtain the leading
small-$q$  behavior of the integrals
\begin{eqnarray}
 I_1^ >  \left( q \right) &=& A_1 \left| q \right| - B_1 \left| q \right|\omega _q  - C_1 q^2  \;,\\
 I_2^ >  \left( q \right) &=& A_2 q^2  - B_2 q^2 \omega _q  + C_2 \left| q \right|^3  \;,\\
 J^ >  \left( q \right)   &=& A_3 \left| q \right| - B_3 q^2\;,
 \label{eq:integrals2}
\end{eqnarray}
where the coefficients generally depend on the cutoff. Note that the constants
$A_1,A_2,A_3,B_1$ and $B_2$ are strictly positive. Using these results, Eqs~(\ref{eq:Static2},\ref{eq:Herring2}) reduce to
\begin{eqnarray}
D_0 &+& A_2 q^2  - \left( {\frac{{K_0 }}{2} - A_1 } \right)\left| q
\right|\phi _q  + I_1^ <  \left( q \right)\phi _q  + I_2^ <  \left(
q \right) = 0\;, \label{eq:Static3} \\
\omega _q &-& \left( {\frac{{K_0 }}{2} - A_3 } \right)\left| q
\right| + J^ <  \left( q \right) = 0 \;.\label{eq:Herring3}
\end{eqnarray}
The advantage of Eqs~(\ref{eq:Static3},\ref{eq:Herring3})
over Eqs~(\ref{eq:Static2},\ref{eq:Herring2}) is that at
the mere price of renormalizing some constants in both equations,
we are left with the integrals $I_1^<(q)$, $I_2^<(q)$ and $J^<(q)$
that can be calculated explicitly for small $q$'s since the power-law form for $\phi_\ell$ and $\omega _\ell$ for small $\ell$'s can
be used. The treatment of Eqs~(\ref{eq:Static3},\ref{eq:Herring3}) is carried on by studying the various
possibilities of balancing the dominant order for small $q$. Note also that the small $q$-dependence of each of the
integrals $I_i^<(q)$ and $J^<(q)$ depend on the convergence of the
integrals without cutoff. So, to leading order in $q$ we get
\begin{eqnarray}
I_1^< (q) &\sim& \left\{
\begin{array}{l}
 E_1|q|\quad  \quad \quad \; 4 - \Gamma  - z > 0 \\
 F_1|q|^{5 - \Gamma  - z}  \quad 4 - \Gamma  - z < 0 \\
 \end{array} \right. \label{eq:integrals3} \\
I_2^<  \left( q \right) &\sim& \left\{ \begin{array}{l}
 E_2 q^2 \quad \quad  \quad \quad 3 - 2\Gamma  - z > 0 \\
 F_2|q|^{5 - 2\Gamma  - z}  \quad 3 - 2\Gamma  - z < 0 \\
 \end{array} \right. \label{eq:integrals4}\\
 J^<  \left( q \right) &\sim& \left\{
\begin{array}{l}
 E_3|q|\quad  \quad \quad \; 4 - \Gamma  - z > 0 \\
 F_3|q|^{5 - \Gamma  - z}  \quad 4 - \Gamma  - z < 0 \\
 \end{array} \right.
\label{eq:integrals5}
\end{eqnarray}

We consider now the quadrant of the $(\Gamma,z)$-plane defined by
$\Gamma>0$ and $z>0$, where solutions may be expected. The lines
$4-\Gamma-z=0$ and $3-2\Gamma-z=0$ divide this quadrant into four
sectors. The classical way \cite{SCE} is to investigate each sector
separately to decide whether or not a solution might exist there.
After performing this analysis we can show that solutions are
possible only in the sector defined by $4 - \Gamma  - z > 0$ and $3
- 2\Gamma - z < 0$, and so we present a detailed analysis for that
sector only, where Eqs~(\ref{eq:Static3},\ref{eq:Herring3}) are
rewritten as
\begin{eqnarray}
D_0  &+& A_2 q^2  - A\left( {\frac{{K_0 }}{2} - A_1  - E_1 }
\right)\left| q \right|^{1 - \Gamma }  - BB_1 \left| q \right|^{1 +
z - \Gamma }  + C_1 \left| q \right|^{2 - \Gamma }  + F_2 q^{5 -
2\Gamma  - z}  = 0 \;,\label{eq:Static5} \\
B|q|^z  &-& \left( {\frac{{K_0 }}{2} - A_3  - E_3 } \right)\left| q
\right| - B_3\left| q \right|^2 = 0\;. \label{eq:Herring4}
\end{eqnarray}
From Eq.~(\ref{eq:Herring4}), it can be easily seen that (in the
small $q$ limit) $z=1$ corresponds to a possible solution. Then
using this result in Eq.~(\ref{eq:Static5}), we find that either
$\Gamma=1$, or $1-\Gamma=5-2\Gamma-z$. The last option implies
$\Gamma = 3$ corresponding to a roughness exponent $\zeta=1$, which
is inconsistent with the assumption of small gradients used to
derive the equation of motion. Therefore, we are left with
$\Gamma=z=1$, which is consistent with the defining condition of
this sector, and is just the solution of the linearized equation of
motion derived from~(\ref{eq:motionFourier}).

Interestingly, a more careful inspection reveals another option
ignored at first sight, namely that of getting a fine-tuned case
where the renormalized nonlocal elasticity vanishes. In
Eq.~(\ref{eq:Static5}) or (\ref{eq:Herring4}), this corresponds to
the case $K_0/2-A_1-E_1=0$ or $K_0/2-A_3-E_3=0$ respectively (note
that the sign of the coefficients is crucial for that argument - and
this is a particular feature of this system which does not happen in
the KPZ system \cite{SCE,KPZ} for example). This fine-tuning
corresponds to being on the critical point.

Repeating the analysis with these options in mind gives rise to
three new possible phases. First, it is possible that the
coefficient of $|q|^{1-\Gamma}$ vanishes in Eq.~(\ref{eq:Static5}),
implying $\Gamma=2$. Then Eq.~(\ref{eq:Herring4}) implies $z=1$ or
$z=2$ according to the coefficient of $|q|$. Another option is when
only the coefficient of $|q|$ vanishes in Eq.~(\ref{eq:Herring4}),
which implies $z=2$, while in Eq.~(\ref{eq:Static5}) as before
$\Gamma=1$. A more careful estimate shows that $A_1+A_3<E_1+E_3$ and
$F_2>0$, and so the only possible cases are with $z=2$. In addition,
since $\Gamma=2$ yields a more singular balance in
Eq.~(\ref{eq:Herring4}) (in the sense that the leading power of $q$
becomes $-1$) compared to the one with $\Gamma=1$, we expect that
this scenario will be the dominant one.

Summarizing this part, we found two possible phases: First, a flat
phase described by $\zeta=0$ and $z=1$, corresponding to the system
in the moving regime. This phase is always possible. Second, we see
the possibility of having a rough phase with $\zeta=1/2$ and $z=2$,
which is possible only on the critical point.

\section{Discussion}
In this paper we analyzed a recently proposed
equation of motion for an in-plane crack front with the aim of
studying possible roughening of the front. We found the possibility
of having a rough moving phase with $\zeta=1/2$ (and $z=2$) which is
relevant for $K_0 \sim K^*$ due to destabilization of the nonlocal
elasticity by the nonlinear term. This result is in agreement with
the roughness exponent measured in experimental systems
\cite{Daguier,Schmittbuhl}. Since in our analysis we neglected the
irreversibility of the fracture process (which becomes important
during the last steps of freezing, and so tends to further roughen
the line), our analysis provides a lower bound for the experimental
results (recall that experimental results vary between
$0.5$--$0.6$). We hope that analysis of the full equation would
yield results which are even closer to the experimental
measurements.

At this point it is useful to comment on what would have happened if
we applied the self-consistent expansion to the full equation of
motion (\ref{eq:motion}) including the KPZ-terms. We checked and
found that the basic analysis is not modified, with the only
difference that the option of having a rough phase with $\zeta=1/2$
and $z=1$ is not ruled out like in the simplified case we discussed
above (due to the fact that some prefactors change). This is
interesting in view of that fact that $z$ has been measured
experimentally in \cite{Schmittbuhl01} and found to be $z=1.2$,
close to the predicted value. This suggests that the KPZ
nonlinearity plays an important role in the dynamics, and cannot be
neglected. Finally, we think that these results are robust in the
sense that they are applicable to other systems having a similar
structure.

\acknowledgments
This work was supported by EEC PatForm Marie Curie
action (E.K.). We thank A. Boudaoud and D. Vandembroucq for fruitful
discussions. Laboratoire de Physique Statistique de l'Ecole Normale
Sup´erieure is associated with the CNRS (UMR 8550) and Universities
Paris VI and Paris VII.

\end{document}